\documentclass{optica-article}

\journal{opticajournal} 

\articletype{Research Article}
\usepackage{float}
\usepackage{pgfplots}
\usepackage{lineno}
\usepackage{booktabs}
\usepackage{setspace}
\usepackage{siunitx}
\newcommand{\SId}[2]{\SI[number-unit-separator=\text{-}]{#1}{#2}}
\DeclareSIUnit{\atperc}{at.\%}
\pgfplotsset{compat=1.18}

\everymath=\expandafter{\the\everymath\displaystyle}

\usepackage[version=4]{mhchem}
\usepackage[nonumberlist,acronym]{glossaries}

\newacronym{klm}{KLM}{Kerr-lens modelocking}
\newacronym{km}{KM}{Kerr medium}
\newacronym{tdl}{TDL}{thin-disk laser}
\newacronym{swir}{SWIR}{short-wave infrared}
\newacronym{xuv}{XUV}{extreme ultraviolet}
\newacronym{cw}{CW}{continuous-wave}
\newacronym{sesam}{SESAM}{semiconductor saturable absorber mirror}
\newacronym{spm}{SPM}{self-phase modulation}
\newacronym{gdd}{GDD}{group-delay dispersion}
\newacronym{roc}{RoC}{radius of curvature}
\newacronym{mlt}{MLT}{Mode-locking telescope}
\newacronym{oc}{OC}{output coupler}
\newacronym{hr}{HR}{highly-reflective}
\newacronym{mpc}{MPC}{multi-pass cell}
\newacronym{ar}{AR}{anti-reflection}
\newacronym{tfp}{TFP}{thin-film polarizer}
\newacronym{fwhm}{FWHM}{full width at half-maximum}
\newacronym{tbp}{TBP}{time-bandwidth product}
\newacronym{rf}{RF}{radio frequency}

\begin{document}

\title{Energy scaling of Kerr-lens modelocked Ho:YAG thin-disk oscillators to the microjoule level}

\author{Sergei Tomilov\authormark{1,*}, Mykyta Redkin\authormark{1}, Yicheng Wang\authormark{1}, Anna Suzuki\authormark{1}, Martin Hoffmann\authormark{1} and Clara J. Saraceno\authormark{1,2}}

\address{\authormark{1}Photonics and Ultrafast Laser Science, Ruhr-Universität Bochum, Universitätstraße 150, 44801 Bochum, Germany\\
\authormark{2}Research Center Chemical Science and Sustainability, University Alliance Ruhr, 44801 Bochum, Germany}

\email{\authormark{*}Sergei.Tomilov@ruhr-uni-bochum.de} 


\begin{abstract*} 
We demonstrate peak power and pulse energy scaling of \gls{klm} \ce{Ho{:}YAG} \glspl{tdl} emitting at \SId{2.1}{\micro\meter} wavelength. We compare different laser configurations to reach a maximum pulse energy of \SI{1.7}{\micro\joule} at an output power of \SI{29}{\watt} with a pulse duration of \SI{434}{\femto\second}, corresponding to a peak power of \SI{3.7}{\mega\watt}. This represents a 5-fold increase in pulse energy and 4-fold increase in peak power compared to previous \gls{klm} \ce{Ho{:}YAG} \glspl{tdl}, thus reaching record high peak power for \SId{2.1}{\micro\meter} modelocked oscillators. We discuss current limitations at this wavelength and guidelines for future work towards higher power and pulse energies.

\end{abstract*}

\section{Introduction}
The \gls{swir} region of the electromagnetic spectrum is currently in growing demand for a variety of applications in science and technology. More specifically, the \SIrange{2}{3}{\micro\metre} wavelength range can be covered by ultrafast lasers directly emitting in this wavelength region, based, for example, on transition-metal (Cr$^{2+}$) and lanthanide (Tm$^{3+}$, Ho$^{3+}$ and Er$^{3+}$) dopants \cite{mirov_frontiers_2018, johnson_optical_2004, fan_spectroscopy_1988, ma_review_2019}, offering enormous potential for extending the performance of modern laser systems to this wavelength region. Such lasers, particularly high-power modelocked oscillators and amplifiers delivering extremely high peak powers and short pulse durations, offer unique advantages for driving the generation of \gls{xuv} pulses \cite{keathley_water-window_2016}, broadband THz transients \cite{clerici_wavelength_2013, wang_efficient_2011} or mid-infrared supercontinua \cite{novak_femtosecond_2018, zhang_multi-mw_2018, nam_octave-spanning_2020} secondary sources; as well as material processing, where SWIR systems are particularly interesting for polymer welding \cite{mingareev_welding_2012} and subsurface modification for silicon processing \cite{richter_sub-surface_2020} among others.

Even though powerful laser systems operating in this wavelength region have seen tremendous progress in the last two decades, their state-of-the-art specifications in terms of average output power, pulse energy, and peak intensity are still behind in comparison with modern 1-µm systems \cite{saraceno_amazing_2019, muller_104_2020, drs_decade_2023, seidel_ultrafast_2024}, with many research efforts to be done. In this context, \glspl{tdl} are well known to be very promising for average power scaling of high repetition rate oscillators \cite{seidel_ultrafast_2024}, but also for future energy scaling of amplifiers \cite{herkommer_ultrafast_2020} in this desired wavelength region – however they have progressed slowly in the \SId{2}{\micro\metre} region, compared, for example, with its fiber laser counterparts based on Tm \cite{gaida_ultrafast_2018, gaida_self-compression_2015}. This is mostly due to the stringent requirements imposed by the disk geometry on the corresponding solid-state gain materials. In fact, the thin-disk geometry imposes not only fabrication constraints to reach high-quality disks (polishing, gluing, etc.) but also operates at high population inversion levels, which in many materials results in detrimental laser operation due to strong quenching. In the thin-disk geometry in \gls{cw} only \SI{24}{\watt} and \SI{5}{\watt}  have been demonstrated with Tm and Cr \cite{zhang_highpower_2018, renz_crznse_2013}. In contrast, Ho-doped YAG has been shown to be better suited for the thin-disk geometry, reaching \SI{50}{\watt} of \gls{cw} output power in multimode configuration \cite{zhang_highpower_2018}. In the first modelocked results, Zhang et al. demonstrated a \SId{25}{\watt} oscillator with \SId{0.325}{\micro\metre} pulse energy and \SId{1.06}{MW} peak power \cite{zhang_kerr-lens_2018}. Further results by our group confirmed that \textgreater\SId{100}{W} level CW single-mode operation is possible \cite{tomilov_towards_2021}. Following these results, we have demonstrated a \ce{Ho{:}YAG}-based \gls{sesam}-modelocked \gls{tdl} single-oscillator with a record output power of \SI{50}{\watt} and a pulse energy of \SI{2}{\micro\joule} \cite{tomilov_50-w_2022}, achieving record-high modelocked power and pulse energy for this wavelength range. The main limitation of this system was a rather long pulse duration of \SI{1.13}{\pico\second} due to the low modulation depth of the saturable absorber. Furthermore, the very novel saturable absorber based on GaSb was not optimized for high power modelocked operation nor shorter pulse durations. While ongoing improvements in \gls{sesam} technology at this wavelength are in progress \cite{heidrich_full_2021}, exploring whether \gls{klm} can provide comparable pulse energies at shorter pulses is also of wide interest to the community, in particular for applications where short pulses are crucial.

In this paper, we demonstrate energy scaling of \gls{klm} \ce{Ho{:}YAG} \glspl{tdl} and present several scalable oscillator designs, including a design with a maximum output energy of \SI{1.7}{\micro\joule} with a pulse duration of \SI{434}{\femto\second} and peak power of \SI{3.7}{\mega\watt}, achieving a 5-fold improvement in
terms of the pulse energy and almost 4-fold in terms of peak power than previously demonstrated
\gls{klm} \glspl{tdl}. To the best of our knowledge, this is the highest peak power demonstrated from any
modelocked oscillator in this wavelength region.

\section{Experimental results}
\subsection{2-reflection resonator design}

\begin{figure}[H]
    \centering
    \begin{spacing}{0.5}
    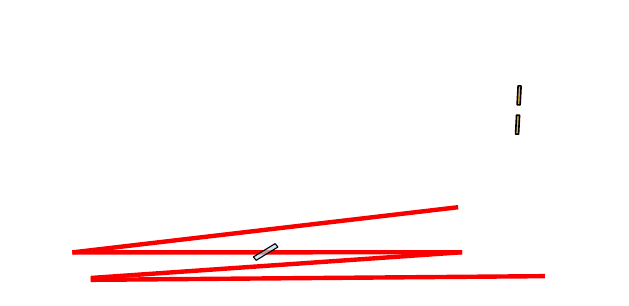
    \end{spacing}
    \caption{Schematic representation of 4-f-extended \gls{klm} cavity.}
    \label{fig:KLM2pass}
\end{figure}

We aim to achieve high average power and pulse energies with a sub-picosecond pulse duration directly from the laser oscillator. This usually requires the usage of high-transmissive \glspl{oc} to operate at moderate intracavity average powers and thus avoid many detrimental thermal effects and nonlinearities \cite{neuhaus_subpicosecond_2008}. In comparison with \ce{Yb{:}YAG}, the most common and robust \SId{1}{\micro\meter} active medium, \ce{Ho{:}YAG} requires operation at reduced doping concentration to avoid deleterious optical transitions. This results in reduced gain per disk reflection and limited loss tolerance compared to Yb-based \glspl{tdl} \cite{tomilov_50-w_2022, barnes_hoho_2003}. To improve the laser efficiency of the resonator containing high-transmissive OC or loss-inducing elements, multiple reflections of the laser mode from the disk are necessary. We start with the easiest configuration possible with two reflections on the disk (i.e. 4 single thickness gain passes) per roundtrip. Fig.~\ref{fig:KLM2pass} illustrates the basic \SId{29}{\mega\hertz} experimental resonator of the KLM TDL. 

It is built around a \gls{mlt}, a short Z-shaped cavity with two 200-mm \gls{roc} mirrors and a \gls{km} between them. To incorporate the disk into the design and implement several mode reflections from the disk, we use 4-f extensions \cite{Schuhmann:18}. The 2-at.\% doped, \SId{190}{\micro\metre} thick \ce{Ho{:}YAG} disk with \SId{2}{\meter} \gls{roc} acts as both mirrors of the 4-f telescope. The dashed line corresponds to the plane where the \gls{mlt} and 4-f extension are connected. Having separate cavity sections responsible for amplification and nonlinear response allows great design modularity for power and energy scaling. The flat end mirror of the \gls{mlt}, 200-mm \gls{roc} mirror next to it, and the \gls{km} are mounted on the linear translation stages, allowing precise tuning of the \gls{mlt} and starting of the modelocking. A total amount of -\SI{24500}{\femto\second\squared} of \gls{gdd} was introduced by flat dispersive mirrors. The oscillator was pumped by a commercial Tm-fiber laser with a maximum power of \SI{209}{\watt} at the wavelength of \SI{1908}{nm}, corresponding to the absorption maximum of \ce{Ho{:}YAG}. The thin-disk multi-pass module enables 72 reflections of the pump beam from the disk, ensuring sufficient absorption of the pump radiation.

A water-cooled copper pinhole was introduced into the beam close to the \gls{oc} position, corresponding to the re-imaged plane marked with the dashed line in Fig.~\ref{fig:KLM2pass}. The hard aperture radius was selected such that the introduced losses almost prevent the system from lasing. However, pushing the end mirror deeper into the stability edge (closer to the nearest curved mirror) reduces clipping of the laser mode on the pinhole, eliminating almost all the losses. 

As part of our optimization procedure, we tested various undoped sapphire and YAG plates of various thicknesses as the \gls{km}. The material and thickness choice influences the strength of the Kerr-lens and introduces \gls{spm} since YAG and sapphire have different $n_2$. Thus, the initial search for modelocked operation was performed with a \SId{4}{\milli\meter} YAG plate and 1-\% OC because high nonlinearity and low intracavity losses allowed for an easier start of modelocking. The \gls{klm} in this system can typically be started with some residual \gls{cw} background, which manifests itself in the form of spikes in the soliton spectrum. The \gls{cw} peaks can be eliminated by increasing the separation of the two curved \gls{mlt} mirrors, pushing the resonator further from the \gls{cw} stability range, and increasing the loss discrimination between \gls{cw} and \gls{klm} modes. Additionally, the pinhole size change allows for the variation of the modelocking mechanism modulation depth, influencing the dampening of \gls{cw} outbreaks and pulse duration.

 \gls{klm} requires to balance intracavity \gls{spm} and \gls{gdd}, therefore both can be used to optimize for stable modelocking. However, the developed design allows more flexibility for nonlinearity tuning due to the movable \gls{km}.  Power scaling of the laser was performed by gradually minimizing intracavity \gls{spm} through plate thickness or $n_2$ reduction and the consequent maximization of intracavity power while having large fixed intracavity \gls{gdd}. As the next step, the \gls{oc} transmissivity was increased to couple a higher fraction of the intracavity power out. The best performance of this system was recorded with a \SId{3}{mm} sapphire plate and a 4-\% \gls{oc}. Thinner plates did not introduce enough self-focusing strength to start and support modelocking, whereas more transparent \glspl{oc} introduced excessive losses, also preventing modelocking. The system emitted up to \SI{14}{\watt} of average power with stable modelocked operation, free of \gls{cw} spikes. This average power corresponds to \SI{350}{\watt} of intracavity power and \SI{0.48}{\micro\joule} of output pulse energy. 

\begin{figure}[H]
    \centering
    \begin{spacing}{0.5}
    \input{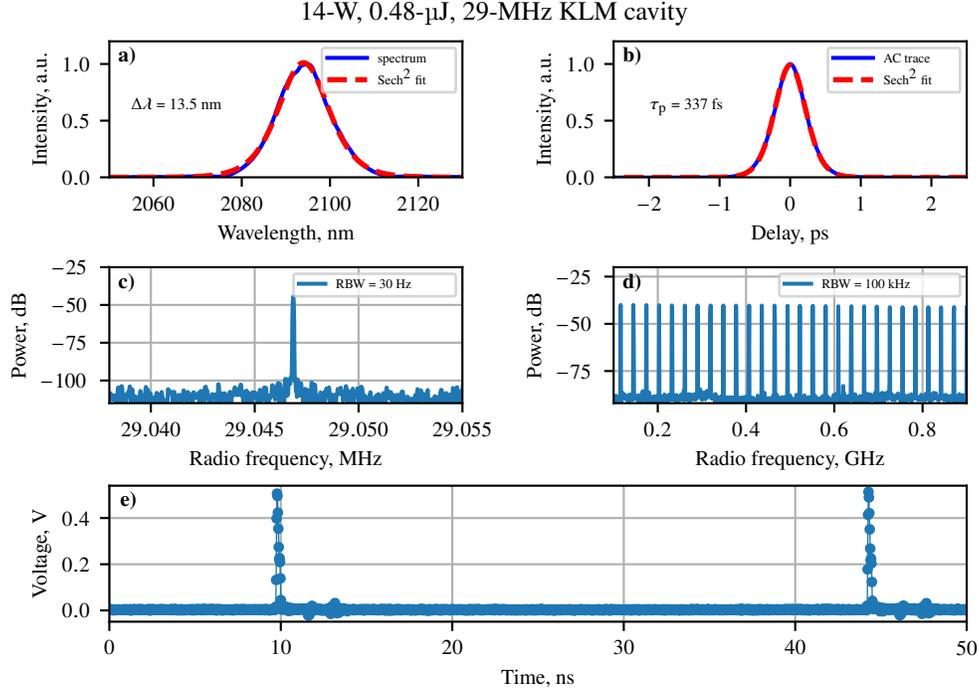}
    \end{spacing}
    \caption{\SId{29}{\mega\hertz} \gls{klm} \gls{tdl} characterization: a) Modelocked laser spectrum, b) autocorrelation trace. Radio-frequency spectra of c) first beatnote, d) harmonics in a \SId{900}{\mega\hertz} span. e) Sampling oscilloscope restoration of one repetition period.}
    \label{fig:shortchar}
\end{figure}

The full characterization data for this laser system is presented in Fig.~\ref{fig:shortchar}. The pulse duration measured with an autocorrelator is \SI{337}{\femto\second} and the spectrum \gls{fwhm} is \SI{13.5}{\nano\meter}. The calculated \gls{tbp} of the pulses is 0.310, slightly below the ideal soliton \gls{tbp} of 0.315, which can be explained by small measurement errors and fitting values. The \gls{rf} pulse train was detected by a fast photodetector with a \SId{10}{\giga\hertz} bandwidth and rise/fall times of \SI{28}{\pico\second}. The Fig.~\ref{fig:shortchar}c) shows the harmonics of this pulse train in a \SId{900}{\mega\hertz} span. The first beatnote is also presented in Fig.~\ref{fig:shortchar}d). It is almost \SI{60}{dB} above the noise level, and its harmonics show almost no decay and/or modulations within the long-span range. This allows us to conclude the absence of Q-switching and short-term instabilities in the signal. Additionally, we used a sampling oscilloscope, which can capture input signals on receiving a trigger event and reconstruct waveforms with a maximum bandwidth of \SI{25}{\giga\hertz}. Using this device and triggering it with the laser repetition frequency, we can sample the pulse train with an effective bandwidth limited by the photodiode. Fig.~\ref{fig:shortchar}e) depicts one sampled repetition period of the studied regime and confirms the perfect fundamentally modelocked operation of the laser. The fluctuations around \SI{13}{\pico\second} and \SI{48}{\pico\second} can be explained by electronic signal interference and do not depend on the laser parameters, such as intracavity average power and pulse bandwidth.

As expected, the \gls{klm} technique allows us to achieve a shorter pulse duration than previously demonstrated by \gls{sesam}-modelocked oscillators \cite{tomilov_50-w_2022}. The double-reflection configuration is capable of modelocking with a maximum \gls{oc} coefficient of 4 \%. In principle, slightly higher values may be possible in future systems with improvements in coating technologies, as we note that dispersive mirrors and disk coatings add significant intracavity loss. In the next section,  we will demonstrate other designs of modelocked oscillators capable of operating with more transparent OCs thanks to the increased number of disk reflections. A further increase of intracavity power through the reduction of the \gls{km}’s thickness proved inefficient with sapphire plates below \SI{3}{\milli\metre} of thickness due to the growing difficulty of starting the modelocking. 
Energy scaling through resonator extension is an alternative approach to improve the laser performance further. This approach would allow us to improve the pulse energy and peak power of the developed laser without introducing significant additional output coupling losses. We therefore explore this path by adding another \SId{4}{\milli\metre} 4-f telescope into the existing cavity to reduce its repetition rate from \SIrange{29}{16.5}{\mega\hertz}. The telescope is based on two spherical \gls{hr} mirrors with a \gls{roc} of \SI{2}{\metre} and is positioned in the plane, marked with the dashed line in Fig.~\ref{fig:KLM2pass}.

\begin{figure}[H]
    \centering
    \begin{spacing}{0.5}
    \input{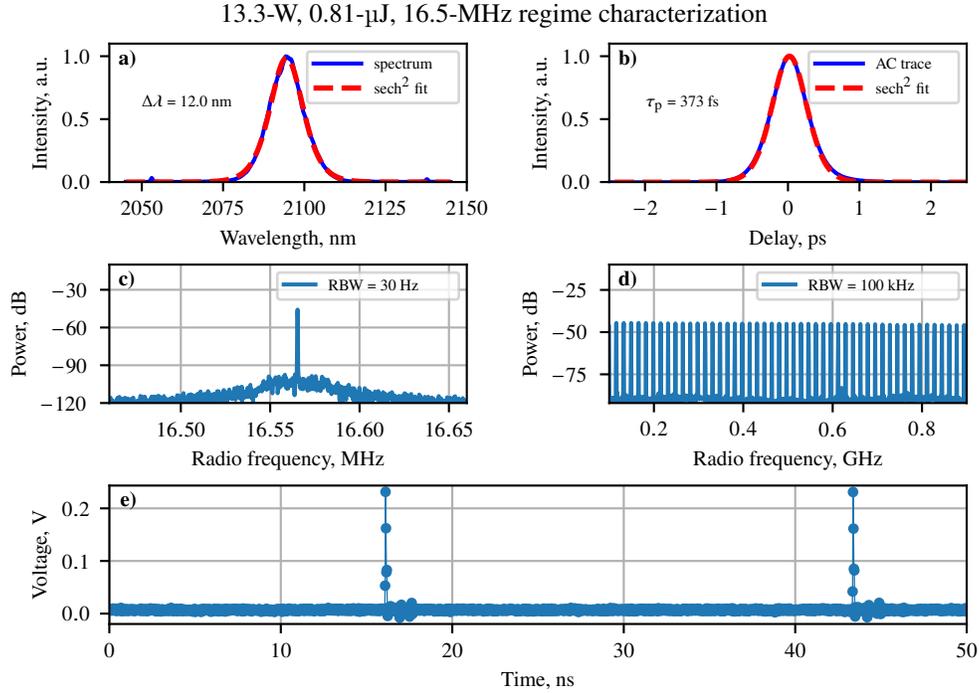}
    \end{spacing}
    \caption{\SId{16.5}{\mega\hertz} \gls{klm} \gls{tdl} characterization: a) Modelocked laser spectrum, b) autocorrelation trace. Radio-frequency spectra of c) first beatnote, d) harmonics in a \SId{900}{\mega\hertz} span. e) Sampling oscilloscope restoration of one repetition period.}
    \label{fig:longchar}
\end{figure}

 We also increased the negative intracavity GDD amount up to \SI{-31500}{\femto\second\squared} to compensate for the rise in \gls{spm} due to increased pulse energy and cavity length. With this cavity, we achieved an output power of \SI{13.3}{\watt} with the 4-\% OC, comparable to the previously demonstrated \SId{29}{\mega\hertz} oscillator. Considering the increased cavity length, this power corresponds now to a pulse energy of \SI{0.8}{\micro\joule}. A pulse duration of \SI{373}{\femto\second} was measured, corresponding to a pulse peak power of \SI{1.9}{\mega\watt}. All data necessary to characterize this performance is presented in Fig.~\ref{fig:longchar}.

\begin{figure}[H]
    \centering

    \input{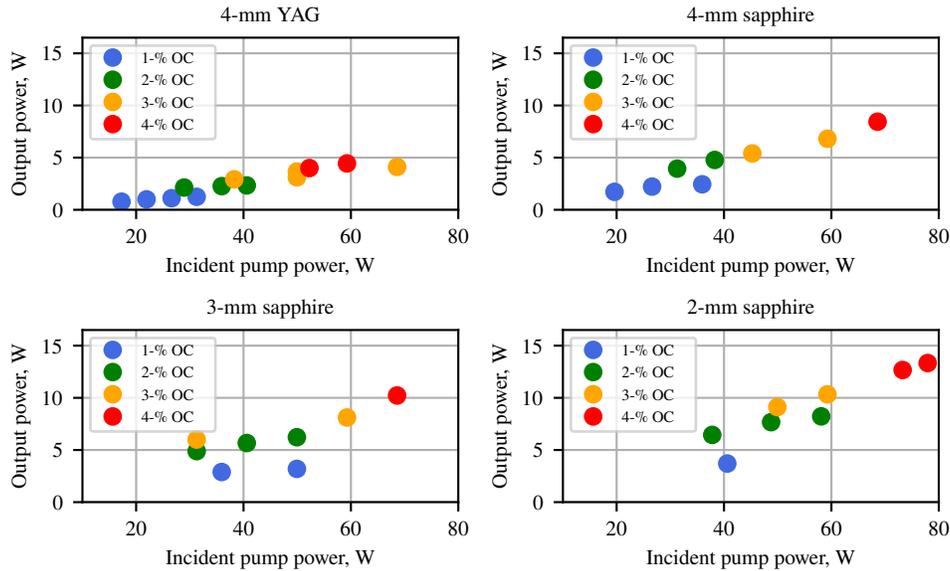}

    \caption{\gls{klm} laser performance as function of \gls{km} thickness, material and \gls{oc} transmissivity}
    \label{fig:plates}
\end{figure}

The characterized performance was achieved with a \SId{2}{mm} sapphire plate as the \gls{km}. Reducing the plate thickness became in this case possible due to the increase in pulse energy. The optimal thickness and the material of the \gls{km} were determined experimentally to support the modelocking with the highest pulse energy and intracavity power. The results of this study are presented in Fig.~\ref{fig:plates}. It highlights the tradeoff between the strength of Kerr lensing and the achievable intracavity power and pulse energy. Plate thickness or $n_2$ reduction improves the system’s performance until the nonlinear self-focusing becomes insufficient to start \gls{klm}. An alternative way to control the intracavity power would be to change the waist of the \gls{mlt} by using mirrors with different curvatures. However, waist reduction will also lead to a growing \gls{spm} contribution from the air, so is also limited in scalability when considering resonators at ambient pressures. In our experiments, we tried MLTs with mirror RoC from \SIrange{100}{300}{\milli\metre}, concluding that a \SIrange{200}{200}{\milli\metre} telescope is optimal for this class of resonators.

The \SId{16.5}{\mega\hertz} resonator was again limited by the maximum tolerable \gls{oc} transmission coefficient of 4 \%. Further scaling of the developed system by optimizing it for higher intracavity powers can be challenging because the further rise in nonlinear phase shift will also require a linear increase of \gls{gdd}. This may lead to an overcomplication of the laser design and the introduction of additional losses and thermal effects. Instead, increasing the oscillator’s loss tolerance, making it operable with higher \gls{oc} transmission coefficients, is the most promising approach. We present first steps in this direction in the following section.

\subsection{3 and 4-reflection \gls{klm} \glspl{km}}

\begin{figure}[H]
    \centering
    \begin{spacing}{0.5}
    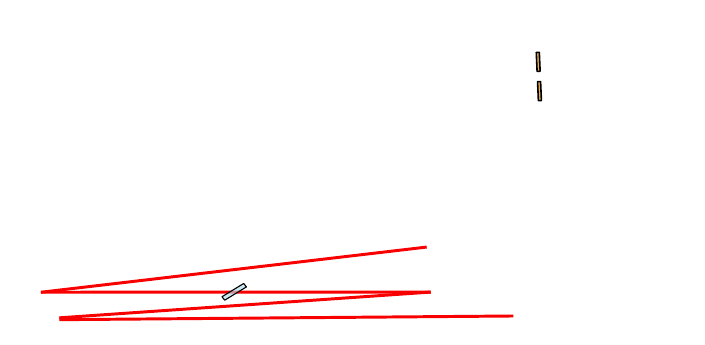
    \end{spacing}
    \caption{\SId{16.7}{\mega\hertz} 3-reflection \ce{Ho{:}YAG} \gls{tdl} oscillator.}
    \label{fig:KLM3pass}
\end{figure}

\begin{figure}[H]
    \centering
    \begin{spacing}{0.5}
    \input{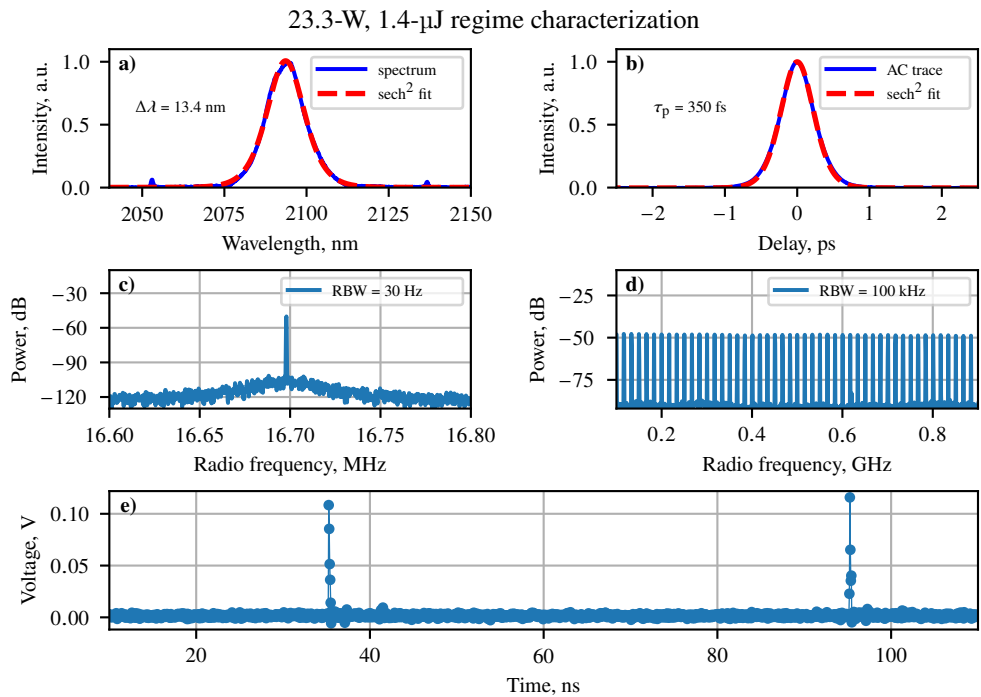}
    \end{spacing}
    \caption{\SId{16.7}{\mega\hertz} 3-reflection \gls{klm} \gls{tdl} characterization: a) Modelocked laser spectrum, b) autocorrelation trace. Radio-frequency spectra of c) first beatnote, d) harmonics in a \SId{900}{\mega\hertz} span. e) Sampling oscilloscope restoration of one repetition period.}
    \label{fig:char3pass}
\end{figure}

The \SId{16.5}{\mega\hertz} oscillator demonstrated in the previous section contains two 4-f extensions formed by concave mirrors and the disk with \SId{2}{\metre} RoC. An increase in loss tolerance for such a design is possible by substituting the curved \gls{hr} mirrors with more reflections of the laser mode on the disk. The first design, which exploits this principle, is shown in Fig.~\ref{fig:KLM3pass}. We were able to remove one of the spherical \gls{hr} mirrors by enabling one more disk reflection and preserving the mode imaging with two 4-f telescopes. We added two 45$^{\circ}$ \gls{hr} mirrors to the design to enable geometrically compact mode folding on the disk, which allowed supporting more than two reflections on it. This common solution is used in \SId{1}{\micro\metre} \gls{tdl} systems featuring multiple disk reflections \cite{zhang_kerr-lens_2018, neuhaus_subpicosecond_2008, saltarelli_power_2019, saltarelli_self-phase_2018}. \gls{klm} operation of this cavity was achieved with \gls{oc} transmission coefficients between 1~\% to 7~\%, reaching \SI{23.3}{\watt} of output power, corresponding to \SI{1.4}{\micro\joule} of pulse energy at a pump power of \SI{162}{\watt} with a 7-\% \gls{oc}. High-transparent \glspl{oc} requires flushing the laser cavity with nitrogen, which eliminates water vapor absorption and improves the laser efficiency. Fig.~\ref{fig:char3pass} depicts the full regime characterization with the maximum achieved output power.
A pulse duration of \SI{350}{\femto\second} was measured, corresponding to \SI{3.5}{\mega\watt} of pulse peak power, featuring an almost 2-fold improvement compared to the aforementioned double-reflection resonator result.

\begin{figure}[H]
    \centering
    \begin{spacing}{0.5}
    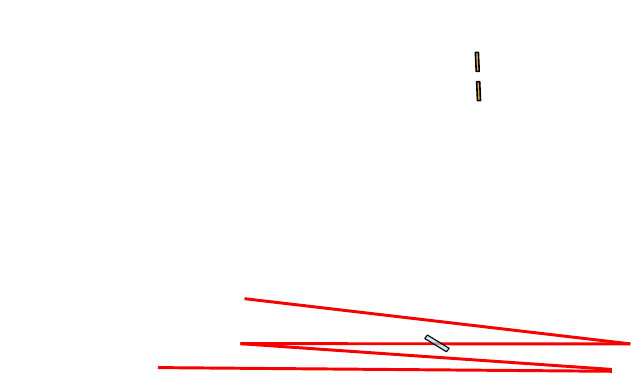
    \end{spacing}
    \caption{\SId{16.5}{\mega\hertz} twin-pump-module \ce{Ho{:}YAG} \gls{tdl} oscillator.}
    \label{fig:Gemini_design}
\end{figure}

Further power scaling of the laser system with higher transmission \glspl{oc} was impossible due to the rapid increase in disk temperature and consequent risk of damage. Developing a \gls{klm} \gls{tdl} requires searching for the stable modelocking conditions by frequently switching between lasing and non-lasing conditions. This leads to an increased thermal and mechanical load on the disk, which temperature approaches 100 $^{\circ}$C at the maximum pump power, potentially resulting in damage. This also points to future developments for Holmium disk systems with thinner disks and more pump passes to minimize operation temperature. To show the possibility of further power scaling and to relax the thermal conditions on the disk, an additional thin-disk pump module with a \SId{2}{\metre} \gls{roc}, 2.5-at.\% doped disk was introduced in the cavity instead of the \SId{2}{\metre} \gls{roc} \gls{hr} mirror as shown in Fig.~\ref{fig:Gemini_design}, and the pump power was split evenly between the two disks. The introduction of the second pump module did not improve the laser slope efficiency.

 On the contrary, introducing the second disk increased passive intracavity losses, illustrating the main problem in the scaling of \ce{Ho{:}YAG} \glspl{tdl} which is loss tolerance sensitivity. However, the improved distribution of pump power allowed us to safely operate the disks at the maximum pump power.

\begin{figure}[H]
    \centering
    \begin{spacing}{0.5}
    \input{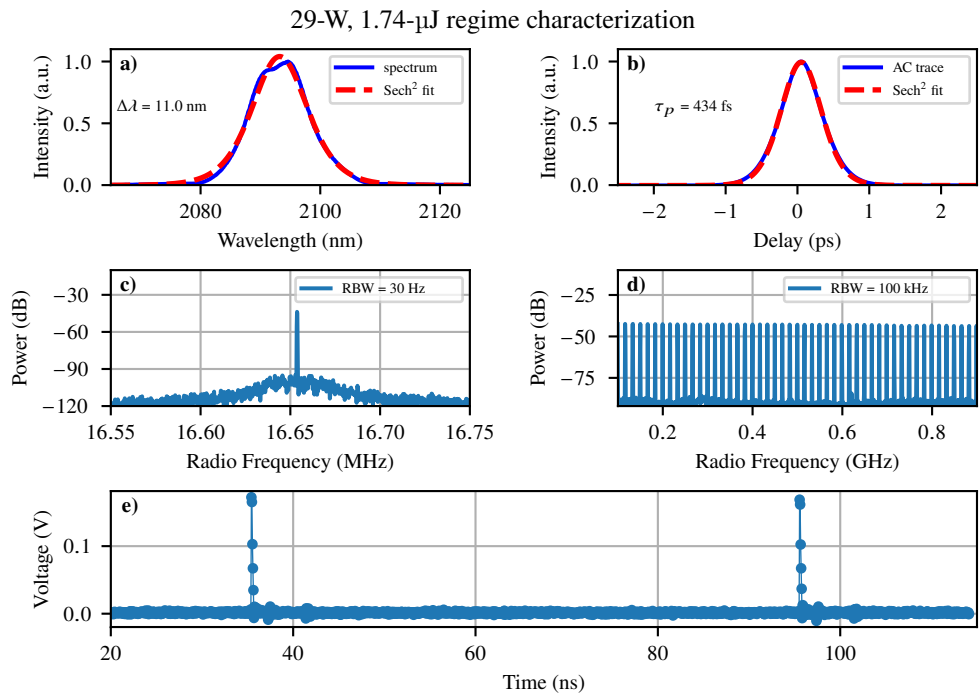}
    \end{spacing}
    \caption{\SId{16.7}{\mega\hertz} twin-pump-module \gls{klm} \gls{tdl} characterization: a) Modelocked laser spectrum, b) autocorrelation trace. Radio-frequency spectra of c) first beatnote, d) harmonics in a \SId{900}{\mega\hertz} span. e) Sampling oscilloscope restoration of one repetition period.}
    \label{fig:gemini_char}
\end{figure}

This allowed us to find stable modelocking conditions within the full pump range with OC transmission coefficients from 1~\% to 10~\%. It is worth noting that high-power ultrafast \SId{1}{\micro\metre} \ce{Yb{:}YAG} \glspl{tdl} are capable of operating with such transparent \glspl{oc} while having only one reflection of the laser mode from the disk \cite{marchese_femtosecond_2008}, again illustrating the unique challenges of Holmium disk technology.

This discrepancy in loss tolerance between \SId{1}{\micro\metre} and \SId{2}{\micro\metre} lasers originates from lower doping concentration and gain cross-sections of \ce{Ho{:}YAG} and requires more careful engineering of the laser resonators based on this medium. As with the previous configurations of \gls{klm} resonators, we fully characterize the stable modelocked laser output of the twin-pump-module laser when using a 10-\% \gls{oc}. The full data is presented in Fig.~\ref{fig:gemini_char}. The system provided a maximum output power of \SI{29}{\watt}, corresponding to a pulse energy of \SI{1.7}{\micro\joule}. A pulse duration of \SI{434}{\femto\second} was measured, resulting in peak power of \SI{3.7}{\mega\watt}. The modelocking stability drastically reduced with an increased OC transmission coefficient. With a 10-\% \gls{oc}, the \gls{klm} start was frequently accompanied by Q-switching spikes, sometimes leading to damage to the \gls{km}. This can be explained by changes in the disk’s gain profile and efficiency reduction from a high population inversion in the active medium. This also results in the more prominent structure of the gain profile, which can explain the distorted spectrum profile at Fig.~\ref{fig:gemini_char}a). The equidistant spacing between output power values achieved with different \glspl{oc} demonstrated in Fig.~\ref{fig:gemini_slopes} allows us to infer that with additional pump power, the demonstrated laser system can operate with \gls{oc} transmissivities of more than 10~\%.

\begin{figure}[H]
    \centering

    \input{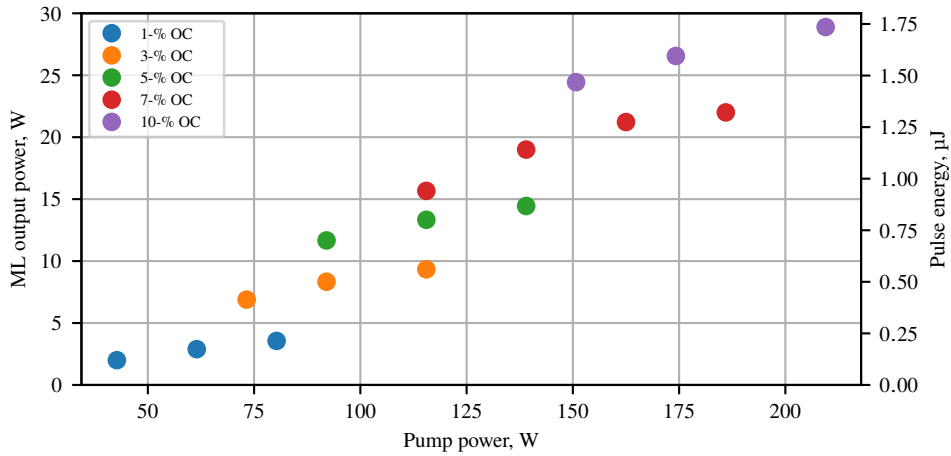}

    \caption{Average power and pulse energy of the twin-pump-module \ce{Ho{:}YAG} \gls{tdl} with different \glspl{oc}}
    \label{fig:gemini_slopes}
\end{figure}

Considering the limited availability of pump sources at \SId{1908}{\nano\metre} wavelength exceeding \SI{200}{\watt} of power and increased heat generation in the thin-disk active medium, adding pump modules with individual disks inside the laser resonator can be an alternative method of power-scaling \SId{2}{\micro\metre} \glspl{tdl}, albeit at increased cost. In the following Section, we will summarize our \gls{klm} results and provide an overview of current limitations and possible ways to mitigate them.

\section{Discussion and outlook}

We summarize the achieved \gls{klm} \gls{tdl} regimes in all cavity configurations in Tab.~\ref{tab:KLM_results}. The twin-pump-module configuration allows us to achieve a maximum pulse energy of \SI{1.7}{\micro\joule} at the maximum available pump power of \SI{209}{\watt}. If we represent these results on the peak power versus pulse energy plot (as in Fig.~10), we can see how \gls{klm} allows us to scale the peak power and shorten the pulse duration in comparison with many ultrafast oscillators and amplifiers in the wavelength range of interest.

\begin{table}[htbp]
    \centering
    \caption{The key parameters of developed \gls{klm} oscillators.}
    \vspace{.3cm}
    \begin{tabular}{clllll}
        \toprule
        Disk reflections & $T_\mathrm{OC}$, \si{\percent} & $P_\mathrm{out}$, \si{\watt} &  $E_\mathrm{p}$, \si{\micro\joule} & $\tau$, \si{\femto\second} & $P_\mathrm{pk}$, \si{\mega\watt}
        \\\midrule
        \num{2} & \num{4} & \num{13.3} & \num{0.8} & \num{373} & \num{1.9} \\
        \num{3} & \num{7} & \num{23.3} & \num{1.4} & \num{350} & \num{3.5} \\
        $\num{3}+\num{1}$ & \num{10} & \num{29} & \num{1.7} & \num{434} & \num{3.7} 
        \\\bottomrule
    \end{tabular}
    \label{tab:KLM_results}
\end{table}

\begin{figure}[H]
    \centering

    \input{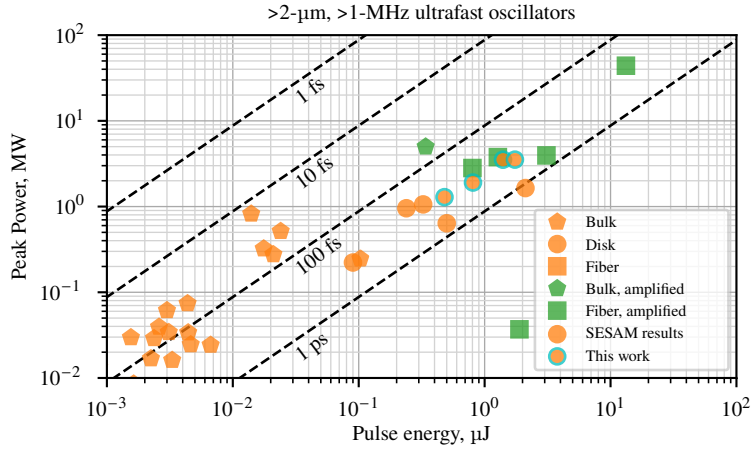}

    \caption{Overview of high-peak-power \gls{swir} oscillators and amplifiers, including achieved \gls{sesam} and \gls{klm} \gls{tdl} results}
    \label{fig:gemini_slopes}
\end{figure}

Hard-aperture \gls{klm} offers flexibility in adjusting the modelocking mechanism’s modulation depth and virtual saturation fluence by choosing the diameter of the aperture and thickness or the material of the \gls{km}. The modulation depth of such a virtual saturable absorber can be controlled by the separation between the \gls{mlt} mirrors or the pinhole diameter, whereas the virtual saturation fluence depends on the reaction of the \gls{km} to the intensity change. For the oscillators demonstrated in this paper, the virtual saturation fluence was estimated to be between \SIlist{10;30}{\micro\joule\per\centi\meter\squared}. These estimations can provide hints for further scaling of the saturation fluence of the \glspl{sesam} developed for \SId{2}{\micro\metre} laser systems.

The main limitation of the \gls{klm} method we encountered experimentally is excessive nonlinearities generated by the \gls{km}, which require significant intracavity GDD to compensate even at low pulse energies. The \SId{1.4}{\micro\joule} 3-pass \gls{klm} \gls{tdl} has an intracavity energy of \SI{20}{\micro\joule} (if we ignore the potential difference in pulse energy throughout the resonator due to leakages and loss/gain sources) and a total nonlinear-phase shift of \SI{1.06}{\radian}. This required \SI{-33500}{\femto\second\squared} of \gls{gdd} for compensation, introduced by 17 flat \SId{-1000}{\femto\second\squared} or \SId{-500}{\femto\second\squared} dispersive mirrors. Dispersive mirrors add additional losses and thermal distortions to the laser mode \cite{saraceno_ultrafast_2014}. In our estimations, up to 20 \% of the total laser output power leaves the resonator not through the \gls{oc} but leaks through the folding dispersive mirrors or
reflects from an imperfectly placed Brewster plate. This sets additional demands for developing new dielectric-coated highly-dispersive mirrors for high-intracavity-power TDLs and highlights
the importance of exceptional quality dielectric top coatings at this wavelength. Furthermore, other techniques for intracavity dispersion compensation can become relevant as the available
tolerance to loss increases. 

In our experiments, the need for larger amounts of GDD explains why, compared to our SESAM modelocking results \cite{tomilov_50-w_2022}, overall modelocked efficiency is lower. However, a full comparison between \gls{klm} and \gls{sesam} modelocking is out of scope of this study and requires a separate thorough investigation; nevertheless, we can safely expect further improvements of both \gls{sesam} and \gls{klm} modelocked \glspl{tdl} in this wavelength regions when larger modulation depth, high-power \glspl{sesam} are demonstrated, and improved low-loss dispersive mirror technology is developed.

Considering pump-power and \gls{oc} limitations, the most natural way of progressing is energy scaling by sacrificing repetition rate. This can be done by implementing longer resonators by stacking 4-f extensions or
introducing an intracavity \gls{mpc} \cite{saraceno_ultrafast_2014} and keeping the average power at the same level. The linear scaling of the energy would, however, also require linear scaling of intracavity \gls{gdd}, so it is beneficial only with coating improvements as mentioned above.

\begin{figure}[H]
    \centering

    \input{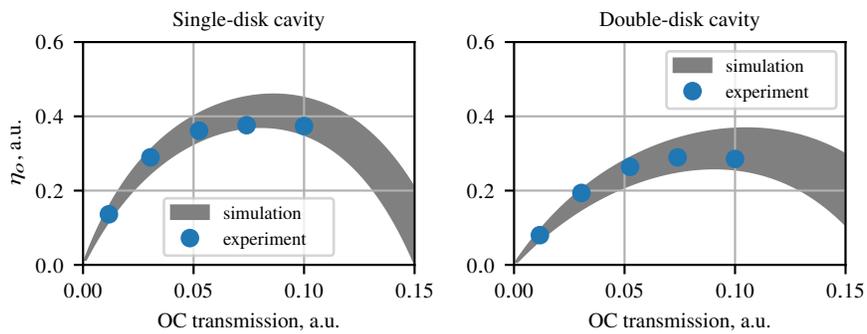}

    \caption{Comparison of experimentally measured slope efficiencies with an analytical model for 3 and 4-disk-reflection laser resonators.}
    \label{fig:slopevsmodel}
\end{figure}

We further explore the importance of passively minimizing intracavity losses by oscillator efficiency analysis. Our experiments with resonators featuring 1 to 4 reflections from the disk in combination with rate-equation models allow us to gather a substantial understanding of the loss tolerance of \ce{Ho{:}YAG} as a thin-disk active medium and determine the optimal amount of disk reflections depending on the \gls{oc} transmissivity. We also measured the \gls{cw} laser slope efficiency without a hard aperture and the \gls{mlt} to exclude all possible sources of non-repeatable passive intracavity losses for the 3-reflection and twin-pump-module resonator configurations. This data was used to perform a Caird analysis \cite{caird_quantum_1988} to determine the passive intracavity losses. This yielded a passive intracavity loss value between 0.054 and 0.071 for 3-reflection and 0.095 and 0.127 for twin-pump-module resonators. This highlights the exceptionally high levels of passive intracavity losses in these cavities. We use the calculated passive loss values to simulate the laser behavior with our analytical models and compare it with the measured slope efficiencies. The result of this comparison is presented in Fig.~\ref{fig:slopevsmodel}. The experimental and theoretical data are in good agreement, considering the imprecision in the results of the Caird analysis, which are mostly caused by variations in the actual OC transmissivity values. This study demonstrates that a twin-pump-module resonator does not improve laser efficiency. On the contrary, due to the almost doubled intracavity passive losses, this design loses around 10~\% of the slope efficiency compared to the 3-reflection oscillator. It also does not provide better loss tolerance when using a 10-\% OC. As mentioned above, the main, if not the only, reason the twin-pump-module oscillator showed better performance is improved thermal management in the active medium.

The mitigation of these thermal issues by improved cooling of the active medium or reduction of the disk thickness can improve the laser performance, thus making the 3-reflection oscillator design an optimum, capable of efficient modelocked operation using any method with \gls{oc} transmissivities up to 10~\%. Using such a resonator for \gls{sesam}-modelocking could also lead to incremental improvements to the results achieved in \cite{tomilov_50-w_2022}. Improved thermal handling of the active medium can also contribute to a higher pump power tolerance of the system. Additionally, one should pay close attention to the minimization of intracavity losses. Aside from the improvement of dielectric coatings discussed above, one can experiment with \gls{ar}-coated \gls{spm} or \gls{km} plates at normal incidence to avoid misalignment and depolarization-induced reflections experienced by a Brewster plate. This would also allow us to arbitrarily define the laser polarization via a \gls{tfp} inside the oscillator and avoid the losses for p-polarization with 45$^\circ$ \gls{hr} mirrors. Placing the system into a vacuum or at least a hermetic nitrogen atmosphere would provide additional improvements for both energy and efficiency scaling since it will minimize \gls{spm} and absorption in the \SId{10}{\metre}-long resonator.

Upon reviewing all these limitations, we can conclude that they can be straightforwardly resolved through engineering efforts. We are convinced that the demonstrated performance of both \gls{sesam} and \gls{klm} Ho:YAG \glspl{tdl} can be scaled further towards \SId{100}{\watt}, \SId{10}{\micro\joule} levels if kW-level pump sources are available and thermal effects as well as pumping efficiency of the disks are properly addressed by optimizing the active medium thickness and the number of pump passes.

\begin{backmatter}
\bmsection{Funding}

\bmsection{Acknowledgment}
We acknowledge support by the DFG Open Access Publication Funds of the Ruhr-Universität Bochum. Funded by the Deutsche Forschungsgemeinschaft (DFG, German Research Foundation) under Germanys Excellence Strategy – EXC-2033 – Projektnummer 390677874 - RESOLV. These results are part of a project that has received funding from the European Research Council (ERC) under the European Union’s HORIZON-ERC-POC programme (Project 101138967 - Giga2u). The research was carried out in the Research Center Chemical Sciences and Sustainability of the University Alliance Ruhr at Ruhr-University Bochum.

\bmsection{Disclosures}
The authors declare no conflicts of interest.

\bmsection{Data availability} No data were generated or analyzed in the presented research.

\end{backmatter}

\bibliography{THESIS_BIB}

\begin{thebibliography}{10}
\newcommand{\enquote}[1]{``#1''}

\bibitem{mirov_frontiers_2018}
S.~B. Mirov, I.~S. Moskalev, S.~Vasilyev, \emph{et~al.}, \enquote{Frontiers of {Mid}-{IR} {Lasers} {Based} on {Transition} {Metal} {Doped} {Chalcogenides},} {\protect\JournalTitle{IEEE J. Select. Topics Quantum Electron.}} \textbf{24}, 1--29 (2018).

\bibitem{johnson_optical_2004}
L.~F. Johnson, \enquote{Optical {Maser} {Characteristics} of {Rare}‐{Earth} {Ions} in {Crystals},} {\protect\JournalTitle{Journal of Applied Physics}} \textbf{34}, 897--909 (2004).

\bibitem{fan_spectroscopy_1988}
T.~Fan, G.~Huber, R.~Byer, and P.~Mitzscherlich, \enquote{Spectroscopy and diode laser-pumped operation of {Tm},{Ho}:{YAG},} {\protect\JournalTitle{IEEE J. Quantum Electron.}} \textbf{24}, 924--933 (1988).

\bibitem{ma_review_2019}
J.~Ma, Z.~Qin, G.~Xie, \emph{et~al.}, \enquote{Review of mid-infrared mode-locked laser sources in the 2.0 \textmu m–3.5 \textmu m spectral region,} {\protect\JournalTitle{Applied Physics Reviews}} \textbf{6}, 021317 (2019).

\bibitem{keathley_water-window_2016}
P.~D. Keathley, G.~J. Stein, P.~Krogen, \emph{et~al.}, \enquote{Water-window soft {X}-ray high-harmonic generation up to the nitrogen {K}-edge driven by a {kHz}, 2.1 µm {OPCPA} source,} in \emph{High-{Brightness} {Sources} and {Light}-{Driven} {Interactions},}  (OSA, Long Beach, California, 2016), p. ET5A.3.

\bibitem{clerici_wavelength_2013}
M.~Clerici, M.~Peccianti, B.~E. Schmidt, \emph{et~al.}, \enquote{Wavelength {Scaling} of {Terahertz} {Generation} by {Gas} {Ionization},} {\protect\JournalTitle{Phys. Rev. Lett.}} \textbf{110}, 253901 (2013).

\bibitem{wang_efficient_2011}
W.-M. Wang, S.~Kawata, Z.-M. Sheng, \emph{et~al.}, \enquote{Efficient terahertz emission by mid-infrared laser pulses from gas targets,} {\protect\JournalTitle{Opt. Lett.}} \textbf{36}, 2608 (2011).

\bibitem{novak_femtosecond_2018}
O.~Novák, P.~R. Krogen, T.~Kroh, \emph{et~al.}, \enquote{Femtosecond {\SI{85}{\micro\meter}} source based on intrapulse difference-frequency generation of {\SI{21}{\micro\meter}} pulses,} {\protect\JournalTitle{Opt. Lett.}} \textbf{43}, 1335 (2018).

\bibitem{zhang_multi-mw_2018}
J.~Zhang, K.~Fai~Mak, N.~Nagl, \emph{et~al.}, \enquote{Multi-{mW}, few-cycle mid-infrared continuum spanning from 500 to 2250 cm$^{-1}$,} {\protect\JournalTitle{Light Sci Appl}} \textbf{7}, 17180--17180 (2018).

\bibitem{nam_octave-spanning_2020}
S.-H. Nam, V.~Fedorov, S.~Mirov, and K.-H. Hong, \enquote{Octave-spanning mid-infrared femtosecond {OPA} in a {ZnGeP} $_{\textrm{2}}$ pumped by a {\SI{24}{\micro\meter}} {Cr}:{ZnSe} chirped-pulse amplifier,} {\protect\JournalTitle{Opt. Express}} \textbf{28}, 32403 (2020).

\bibitem{mingareev_welding_2012}
I.~Mingareev, F.~Weirauch, A.~Olowinsky, \emph{et~al.}, \enquote{Welding of polymers using a {\SI{2}{\micro\meter}} thulium fiber laser,} {\protect\JournalTitle{Optics \& Laser Technology}} \textbf{44}, 2095--2099 (2012).

\bibitem{richter_sub-surface_2020}
R.~A. Richter, N.~Tolstik, S.~Rigaud, \emph{et~al.}, \enquote{Sub-surface modifications in silicon with ultra-short pulsed lasers above 2 µm,} {\protect\JournalTitle{J. Opt. Soc. Am. B}} \textbf{37}, 2543 (2020).

\bibitem{saraceno_amazing_2019}
C.~J. Saraceno, D.~Sutter, T.~Metzger, and M.~Abdou~Ahmed, \enquote{The amazing progress of high-power ultrafast thin-disk lasers,} {\protect\JournalTitle{J. Eur. Opt. Soc.-Rapid Publ.}} \textbf{15}, 15 (2019).

\bibitem{muller_104_2020}
M.~Müller, C.~Aleshire, A.~Klenke, \emph{et~al.}, \enquote{10.4 {kW} coherently combined ultrafast fiber laser,} {\protect\JournalTitle{Opt. Lett.}} \textbf{45}, 3083--3086 (2020).

\bibitem{drs_decade_2023}
J.~Drs, J.~Fischer, N.~Modsching, \emph{et~al.}, \enquote{A {Decade} of {Sub}-100-fs {Thin}-{Disk} {Laser} {Oscillators},} {\protect\JournalTitle{Laser \& Photonics Reviews}} \textbf{17}, 2200258 (2023).

\bibitem{seidel_ultrafast_2024}
M.~Seidel, L.~Lang, C.~R. Phillips, and U.~Keller, \enquote{Ultrafast 550-{W} average-power thin-disk laser oscillator,} {\protect\JournalTitle{Optica}} \textbf{11}, 1368 (2024).

\bibitem{herkommer_ultrafast_2020}
C.~Herkommer, P.~Krötz, R.~Jung, \emph{et~al.}, \enquote{Ultrafast thin-disk multipass amplifier with 720 {mJ} operating at kilohertz repetition rate for applications in atmospheric research,} {\protect\JournalTitle{Opt. Express}} \textbf{28}, 30164--30173 (2020).

\bibitem{gaida_ultrafast_2018}
C.~Gaida, M.~Gebhardt, T.~Heuermann, \emph{et~al.}, \enquote{Ultrafast thulium fiber laser system emitting more than 1 {kW} of average power,} {\protect\JournalTitle{Opt. Lett.}} \textbf{43}, 5853 (2018).

\bibitem{gaida_self-compression_2015}
C.~Gaida, M.~Gebhardt, F.~Stutzki, \emph{et~al.}, \enquote{Self-compression in a solid fiber to 24 {MW} peak power with few-cycle pulses at 2 \textmu m wavelength,} {\protect\JournalTitle{Opt. Lett.}} \textbf{40}, 5160 (2015).

\bibitem{zhang_highpower_2018}
J.~Zhang, F.~Schulze, K.~F. Mak, \emph{et~al.}, \enquote{High‐{Power}, {High}‐{Efficiency} {Tm}:{YAG} and {Ho}:{YAG} {Thin}‐{Disk} {Lasers},} {\protect\JournalTitle{Laser \& Photonics Reviews}} \textbf{12}, 1700273 (2018).

\bibitem{renz_crznse_2013}
G.~Renz, J.~Speiser, A.~Giesen, \emph{et~al.}, \enquote{Cr:{ZnSe} thin disk cw laser,}  (San Francisco, California, USA, 2013), p. 85991M.

\bibitem{zhang_kerr-lens_2018}
J.~Zhang, K.~F. Mak, and O.~Pronin, \enquote{Kerr-{Lens} {Mode}-{Locked} {\SId{2}{\micro\meter}} {Thin}-{Disk} {Lasers},} {\protect\JournalTitle{IEEE J. Select. Topics Quantum Electron.}} \textbf{24}, 1--11 (2018).

\bibitem{tomilov_towards_2021}
S.~Tomilov, M.~Hoffmann, Y.~Wang, and C.~J. Saraceno, \enquote{Moving towards high-power thin-disk lasers in the 2 \(\mu\)m wavelength range,} {\protect\JournalTitle{Journal of physics: Photonics}} \textbf{3}, 022002--1 -- 022002--12 (2021).

\bibitem{tomilov_50-w_2022}
S.~Tomilov, Y.~Wang, M.~Hoffmann, \emph{et~al.}, \enquote{50-{W} average power {Ho}:{YAG} {SESAM}-modelocked thin-disk oscillator at 2.1µm,} {\protect\JournalTitle{Opt. Express, OE}} \textbf{30}, 27662--27673 (2022).

\bibitem{heidrich_full_2021}
J.~Heidrich, M.~Gaulke, B.~O. Alaydin, \emph{et~al.}, \enquote{Full optical {SESAM} characterization methods in the 19 to 3-µm wavelength regime,} {\protect\JournalTitle{Opt. Express}} \textbf{29}, 6647 (2021).

\bibitem{neuhaus_subpicosecond_2008}
J.~Neuhaus, D.~Bauer, J.~Zhang, \emph{et~al.}, \enquote{Subpicosecond thin-disk laser oscillator with pulse energies of up to 259 microjoules by use of an active multipass geometry,} {\protect\JournalTitle{Opt. Express}} \textbf{16}, 20530 (2008).

\bibitem{barnes_hoho_2003}
N.~P. Barnes, B.~M. Walsh, and E.~D. Filer, \enquote{Ho:{Ho} upconversion: applications to {Ho} lasers,} {\protect\JournalTitle{J. Opt. Soc. Am. B}} \textbf{20}, 1212 (2003).

\bibitem{Schuhmann:18}
K.~Schuhmann, K.~Kirch, M.~Marszalek, \emph{et~al.}, \enquote{Multipass amplifiers with self-compensation of the thermal lens,} {\protect\JournalTitle{Appl. Opt.}} \textbf{57}, 10323--10333 (2018).

\bibitem{saltarelli_power_2019}
F.~Saltarelli, I.~J. Graumann, L.~Lang, \emph{et~al.}, \enquote{Power scaling of ultrafast oscillators: 350-{W} average-power sub-picosecond thin-disk laser,} {\protect\JournalTitle{Opt. Express}} \textbf{27}, 31465 (2019).

\bibitem{saltarelli_self-phase_2018}
F.~Saltarelli, A.~Diebold, I.~J. Graumann, \emph{et~al.}, \enquote{Self-phase modulation cancellation in a high-power ultrafast thin-disk laser oscillator,} {\protect\JournalTitle{Optica}} \textbf{5}, 1603 (2018).

\bibitem{marchese_femtosecond_2008}
S.~V. Marchese, C.~R. Baer, A.~G. Engqvist, \emph{et~al.}, \enquote{Femtosecond thin disk laser oscillator with pulse energy beyond the 10-microjoule level,} {\protect\JournalTitle{Opt. Express}} \textbf{16}, 6397 (2008).

\bibitem{saraceno_ultrafast_2014}
C.~J. Saraceno, F.~Emaury, C.~Schriber, \emph{et~al.}, \enquote{Ultrafast thin-disk laser with {\SI{80}{\micro\joule}} pulse energy and 242 {W} of average power,} {\protect\JournalTitle{Opt. Lett.}} \textbf{39}, 9 (2014).

\bibitem{caird_quantum_1988}
J.~Caird, S.~Payne, P.~Staber, \emph{et~al.}, \enquote{Quantum electronic properties of the {Na$_3$}{Ga$_2$}{Li$_3$}{F$_{12}$}:{Cr$^{3+}$} laser,} {\protect\JournalTitle{IEEE J. Quantum Electron.}} \textbf{24}, 1077--1099 (1988).

\end{thebibliography}






\end{document}